# Diamond Integrated Optomechanical Circuits

Patrik Rath[1], Svetlana Khasminskaya[1], Christoph Nebel[2], Christoph Wild[2,3] and Wolfram H.P. Pernice[1,*]

[1] *Institute of Nanotechnology, Karlsruhe Institute of Technology, Hermann-von-Helmholtz-Platz 1, 76344 Eggenstein-Leopoldshafen, Germany*
[2] *Fraunhofer Institute for Applied Solid State Physics, Tullastr. 72, 79108 Freiburg, Germany*
[3] *Diamond Materials GmbH, Tullastr. 72, 79108 Freiburg, Germany*

[*] *Corresponding author electronic mail: wolfram.pernice@kit.edu*

**Diamond offers unique material advantages for the realization of micro- and nanomechanical resonators due to its high Young's modulus, compatibility with harsh environments and superior thermal properties. At the same time, the wide electronic bandgap of 5.45eV makes diamond a suitable material for integrated optics because of broadband transparency and the absence of free-carrier absorption commonly encountered in silicon photonics. Here we take advantage of both to engineer full-scale optomechanical circuits in diamond thin films. We show that polycrystalline diamond films fabricated by chemical vapour deposition provide a convenient waferscale substrate for the realization of high quality nanophotonic devices. Using free-standing nanomechanical resonators embedded in on-chip Mach-Zehnder interferometers, we demonstrate efficient optomechanical transduction via gradient optical forces. Fabricated diamond resonators reproducibly show high mechanical quality factors up to 11,200. Our low cost, wideband, carrier-free photonic circuits hold promise for all-optical sensing and optomechanical signal processing at ultra-high frequencies.**

# Introduction

Photonic integrated circuits offer excellent prospects for high speed, complex optical applications on a chip scale. Relying on established fabrication procedures originally developed for the micro-electronics industry, high quality, sub-wavelength devices can be manufactured as the optical analogue of electronic integrated circuits[1-3]. Because nanophotonic devices are size-matched to nanoscale mechanical resonators[4,5], the combination of these formerly separate fields into optomechanical circuits provides additional degrees of freedom for tunable optical components[6,7]. In recent years the investigation of integrated optomechanical circuits has therefore received considerable attention for the investigation of fundamental physics[8-11], as well as for applications in metrology[12,13] and sensing[14]. Silicon-on-insulator (SOI) has emerged as one of the most promising platforms for such devices. However, due to a limited bandgap of 1.1eV silicon is only transparent for wavelengths above 1100nm, excluding the entire visible wavelength range which is of major importance for applications in fluorescent imaging and biology. Therefore alternative material systems with a wider bandgap, such as silicon nitride[15-17], aluminum nitride[18,19] or gallium nitride[20,21] are attracting increasing interest in order to enlarge the wavelength operation range of nanophotonic devices. In addition, novel photonic materials such as diamond[22,23] offer further possibilities by combining a large bandgap of 5.45eV with attractive material properties, such us high thermal conductivity, chemical inertness and biocompatibility. In addition, diamond possesses a high refractive index of 2.4[24], which allows for efficient confinement of light in nanoscale waveguides when surrounded by suitable cladding materials. Furthermore, diamond also offers one of the largest Young's moduli found in nature, about 5-6 larger than silicon's, which provides a route for the fabrication of high-frequency mechanical devices[25]. Previous studies in nanocrystalline diamond demonstrated nanomechanical resonators with frequencies up to 640MHz[26,27]. In this case the mechanical material properties proved to be comparable to the ones found in bulk diamond, suggesting that polycrystalline diamond may be a suitable substitute for the realization of on-chip nanomechanical resonators.

To date the use of diamond for optical applications has predominantly been restricted to single-crystalline substrates, which can be grown with high purity using chemical vapour deposition (CVD). This inherently limits the size of such substrates for integrated photonics and also requires sophisticated transfer techniques to prepare diamond-on-insulator (DOI)[22] substrates, conceptually similar to SOI. Here in contrast we show that polycrystalline diamond thin films directly deposited onto large substrates provide a wafer-scale template for the fabrication of high-quality integrated optical circuits. Even without surface polishing, high optical transmission at telecoms wavelengths can be achieved. Using partially etched ridge waveguides we obtain propagation loss of 53dB/cm, limited by scattering due to the residual surface roughness. Optical waveguiding over long-distances on-chip provides the ingredients for photonic networks and allows for the realization of functional optomechanical devices. We fabricate free-standing nanomechanical resonators which are actuated efficiently via gradient optical forces[4,5], strong enough to lead to both stiffening and softening Duffing non-linearity, depending on the residual internal stress. The microcrystalline diamond used in our study can be grown on substrates with up to 6 inch in diameter with low levels of impurities[28,29] and allows for high mechanical quality factors without further surface polishing treatment. Our demonstration of diamond integrated optomechanical circuits opens new routes towards an all-optically driven platform for fundamental science and advanced sensing applications.

# Results

**Microcrystalline Diamond-on-Insulator nanophotonic circuits**

In order to enable optical waveguiding diamond needs to be surrounded by a buffer or cladding material with lower refractive index. Previously this was achieved by transferring diamond thin films to carrier wafers with a silicon dioxide cladding layer[22,23]. This requires advanced bonding techniques and subsequent etching or polishing steps in order to reduce the diamond thickness to target layer thickness of less than a micron. While this approach is sufficient for the fabrication of small-area devices, it does not allow for the fabrication of functional nanophotonic circuits commonly employed in integrated optics. Furthermore, during chemo-mechanical polishing uniform layer thickness is especially hard to control with diamond substrates, leading to surface variations and thickness non-uniformity[21].

In this work these limitations are overcome by direct overgrowth of wafer-scale substrates with diameters up to 6 inch. We employ high-quality silicon substrates, which provide an atomically flat surface. A low-refractive index buffer layer is grown by thermally oxidizing the wafer to a thickness of 2µm. During the oxidation process the surface morphology is preserved, providing a smooth starting layer for the later diamond growth. Microcrystalline diamond layers are then deposited directly onto the oxidized substrates. To initiate the later growth of diamond on our substrates, firstly, a diamond nano-particle seed layer is deposited onto the $SiO_2$ film by ultrasonification for 30 minutes in a water based suspension of ultra-dispersed (0.1 wt %) nano-diamond particles of typically 5-10nm size[30]. Then samples are rinsed with de-ionized water and methanol. After dry blowing, the wafer is transferred into an ellipsoidal 915MHz microwave plasma reactor[31] where diamond films with a thickness below 1µm are grown at 1.8kW microwave power, using 2% $CH_4$ in 98% $H_2$, at a pressure of 80mbar and a temperature of 850°C. Substrate rotation was applied to avoid angular non-uniformities arising from the gas flow. Growth rates were in the range of 1-2µm/h. After growth the samples are cleaned in concentrated $HNO_3$:$H_2SO_4$ to remove surface contaminations.

Diamond nanophotonic circuits are then realized using a combination of electron-beam (e-beam) lithography and pattern transfer into the thin film via reactive ion etching. We employ the negative e-beam resist Fox15, which enables resist thickness of several 100 nanometers. Upon exposure Fox15 provides an anorganic glass-like protection layer, which is resistant to oxygen plasma in contrast to organic e-beam resists. High-resolution photonic circuits are written using a JEOL 5300 50kV e-beam system. After developing, the negative tone images are transferred into the diamond layer using reactive ion etching in an Oxford 80 Plasmalab etcher in oxygen-argon ($O_2$/Ar) chemistry. The etching step is carefully controlled by timing in order to obtain a desired target depth within the diamond substrate.

In order to realize free-standing waveguides as shown schematically in Fig.1a), the oxide layer underneath the diamond needs to be removed. Therefore a second e-beam lithography step is performed in positive-tone PMMA resist. Because PMMA does not withstand the RIE procedure necessary to etch diamond, the wafer is first coated with a thin layer of chromium using e-beam evaporation. Opening windows across desired sections of the waveguides are defined in PMMA and the chromium in the open areas is removed using Cr-etchant in a first wet etching step. Then the now exposed diamond in the open areas is reactively etched in $O_2$/Ar-chemistry all the way through, thus revealing the underlying substrate. The as-defined windows are then isotropically etched in buffered oxide etchant (BOE) to a depth of roughly 1µm. Here we note that the two-step etching routine allows us to precisely define the mechanical resonator length and provides controlled clamping points for the beams. This fabrication flow is depicted in Fig.1b), further details on our fabrication procedure are provided in the Methods section. A cross-sectional scanning electron microscope (SEM) image of a resulting waveguide is presented in Fig.1c). We show the horizontal cut through a fabricated waveguide, still covered with the remaining HSQ e-beam resist. The waveguide is a ridge waveguide, etched partially into the microcrystalline diamond layer. As apparent from the image the RIE process leads to near-vertical sidewalls. Also revealed is the underlying microcrystalline structure of the CVD diamond with a typical grain size on the order of 100nm.

We fabricate optical waveguides with a width of 1µm, etched 300nm into the diamond layer in order to obtain efficient guiding of light at 1550nm target wavelength. In order to characterize the transmission properties of as-fabricated devices we use transmission measurements in the telecoms C- and L-bands. To couple light into the waveguides we design compact focusing grating couplers as shown in the SEM image in the inset of Fig.1d). Light from a continuously tunable laser source (New Focus TLB-6600) is coupled into the fabricated chips using an optical fiber-array aligned to input and output grating couplers. Transmitted light is recorded with a low-noise photodetector (New Focus 2011). Optimized couplers lead to input coupling loss of 5dB, in par with previously reported results on silicon and AlN substrates[4,17,18]. The grating couplers provide a 3dB-coupling bandwidth of roughly 50nm, which can be tuned across the C- and L-bands by varying the period of the grating as shown in Fig.1d). By measuring waveguides with varying length we can then extract the on chip propagation loss. We fabricate waveguides with a length up to 4.6mm, from which we obtain propagation loss $\alpha$ of 52.8 ± 4.2dB/cm. Corresponding propagation loss is also independently measured by fabricating microring resonators with a radius of 40µm. In such devices we obtain optical quality factors up to 8,500 by fitting the resonance dip with a Lorentzian function. Then the propagation loss can be extracted as $\alpha = 10\log_{10} e \cdot 2\pi n_g / Q_{int}\lambda$ (where $Q_{int}$ is the intrinsic quality factor, $\lambda$ the wavelength and $n_g$ is the group index), which yields a loss value of 51dB/cm, in good agreement with the loss measured from long waveguides. This propagation loss is mainly due to scattering events occurring at the surface roughness of the as-grown diamond layer. We employ atomic force microscopy (AFM) to measure the topology of the diamond after CVD growth. From the data (see Methods section), we determine a root-mean square (rms) roughness of 15.4nm rms. Using the rms roughness value an upper bound on the propagation loss can be estimated using the Payne-Laycey (PL) model[32,33]. For our waveguide geometry the estimated scattering loss due to surface non-uniformity amounts to 55-96dB/cm, which is on the same order as our measured value. This is also consistent with the fact, that we do not find a significant amount of non-diamond bound carbon (which would lead to optical absorption at grain boundaries), as estimated from Raman spectroscopy (see Methods section). Thus the propagation loss could be further reduced by employing polishing techniques to remove the surface roughness. However, for the investigation of optomechanical circuits with typical waveguide lengths on the order of a few hundred microns, the residual propagation loss can easily be tolerated.

**Interferometric detection of nanomechanical motion**

In order to characterize the optomechanical response of nanomechanical resonators embedded in our on-chip circuits, phase-sensitive measurements have to be performed. For this purpose we design integrated Mach-Zehnder interferometers as depicted in Fig.2a). We realize hundreds of devices on each chip in order to scan relevant optomechanical parameters. Both arms of the interferometer contain mechanical devices in order to provide balanced interference at the output Y-splitter. A SEM image of the released section of one of the MZIs is shown in Fig.2b). In order to provide significant optical forces on chip for the actuation of the mechanical resonators, waveguide geometries that lead to strong field gradients in the optical mode are preferred. In dielectric waveguides strong gradients can be obtained by employing slot-geometries, which consist of two dielectric regions separated by a narrow air gap[34]. When released from the underlying substrate, slot waveguides form laterally coupled nano-mechanical resonators which provide a convenient geometry to tailor opto-mechanical interactions[5]. Such slot waveguides are realized in our devices by splitting the input waveguide of each MZI-arm in two over a limited distance of 100μm in the straight section of the interferometer. By adiabatically tapering the transition from a single-waveguide to the slot waveguide scattering loss is avoided, thus maintaining high transmission through the device. Resulting field distributions for the unreleased slot waveguide and the free-standing parallel diamond beams are shown in the simulated profile of the electric field distribution in Fig.2c) and Fig.2d), respectively, with the strong field confinement clearly visible for the released beams. By releasing the slotted section of the MZI two pairs of mechanical resonators are therefore created in each nanophotonic circuits. The largest field gradients are obtained when narrow air gaps are used. Therefore we vary the distance between in the dielectric waveguide sections between 150nm and 300nm on different devices. We also vary the width of the dielectric waveguide beams between 350nm and 500nm in order to modify the stiffness of the released mechanical devices while making sure that the desired optical mode is guided in the slot waveguide.

The spectral performance of the MZI-devices is characterized by scanning the tunable laser in continuous swept mode across the coupler bandwidth. The measured transmission through the on-chip MZI is shown in Fig.2e) where we find the expected interference fringes with a free-spectral range of 9.5nm. The device is well balanced, thus providing high extinction ratio of up to 25dB as shown in the logarithmic plot in the inset of Fig.2e). The high visibility of the on-chip MZI enables sensitive detection of phase variations occurring in either of the two arms. The mechanical vibration of the released waveguide sections causes an induced phase change by modifying the effective refractive index of a propagating mode. When the beams move closer together, the effective index increases, thus increasing the phase difference between light propagating through the interferometer arms. By positioning a probe laser beam on the quadrature points of the MZI we are therefore able to resolve the mechanical motion of the released beams in the transmitted optical signal. In order to identify the mechanical resonances we record the transmission with a fast photodetector (New Focus 1811) and analyze the response with a network/spectrum analyzer (R&S ZVL6). To reduce air damping, the optomechanical circuits are mounted in a vacuum chamber with a base pressure of $10^{-6}$mTorr. Typical results for the thermomechanical motion of selected devices are shown in Fig.3a). Here we plot the resonance amplitude for devices with a length varying from 10µm to 25µm, with decreasing length from the top left to the bottom right spectrum. The exact dimensions of each device are indicated within the sub-panels of Fig.3a). By fitting the measured power spectral density (PSD) with a Lorentzian function we can extract the mechanical quality factor (Q-factor) of the devices. We can calibrate the amplitude of the Brownian motion of the different beams by comparing the measured spectral density with the expected spectral density of the displacement noise at resonance frequency $S_y = \sqrt{4 k_B T\, Q / m_{eff} (2\pi f)^3}$ as known for the calibration of AFM cantilevers, with the Boltzmann constant $k_B$, the temperature T = 300K and the mechanical quality factor $Q$. Using the volume of the beam $V$, its physical mass $m_0$, the displacement $u(x, y, z)$ and the maximum displacement $u_{max}$ we can calculate the effective modal mass as $m_{eff} = \frac{m_0}{u^2_{max} \cdot V} \iiint u^2(x, y, z) \, dx \, dy \, dz$.[35] The measured Q-factors in dependence of resonance frequency are plotted in Fig.3b). A dramatic increase for lower frequency devices is observed, consistent with previously reported results. For the highest frequency devices with a length of 10µm and a beam width of 350nm we obtain Q-factors on the order of 950. When coupling higher optical power into the waveguides, good displacement sensitivity can be reached with this waveguide geometry[4]. Taking into account the coupling loss occurring at the input of the devices, the propagation loss to the released section of the beam as well as the power loss due to the input splitter, we estimate optical power on the beam of 1.6mW. In this case the measured displacement sensitivity amounts to $11 \times 10^{-15}$ mHz$^{-1/2}$, obtained from the baseline of the thermomechanical noise spectra. The best mechanical Q-factor is measured as 11,200 at 3.8MHz (the corresponding damping rates f/Q are shown as the blue markers on the right axis of Fig.3b)), exceeding the values obtained for nanocrystalline diamond[26] and crystalline silicon nanomechanical resonators of comparable dimensions[4,36]. Higher mechanical quality factors have recently been reported for single crystalline diamond cantilevers[24], with dimensions on a tens of micrometer scale at lower frequencies. We therefore expect that by moving towards longer nanomechanical resonators the mechanical Q-factors in our waveguide resonators may also be further improved. By varying the beam cross-section of our waveguides we also cover higher resonance frequencies as reported in reference 24, in our case from 2MHz up to 40MHz as shown in Fig.3c). The expected inverse quadratic dependence on resonator length is observed, as well as a linear increase in resonance frequency when the beam width is increased.

**Gradient force actuation of beam waveguides**

Because the slot-waveguide geometry leads to tight confinement of light into the gap between neighboring waveguides, significant gradient optical forces can be generated[5]. Thus in-plane mechanical modes as shown in the simulation image in the inset of Fig.3c) can be excited all-optically without the need for electrical connections to the chip. In order to demonstrate optomechanical actuation we therefore employ a pump-probe measurement scheme[4] as depicted in Fig.4a). The setup allows us to measure the driven response of the nanomechanical beams. A weak probe laser is positioned on one of the slopes of the interference fringes as shown in the bottom left inset of Fig.4a). A second pump laser is placed in one of the dips of the MZI-transmission, far enough away from the probe wavelength to provide direct optical attenuation of the transmitted signal by 20dB. The pump laser is amplitude modulated with a fast lithium niobate electro-optical modulator (EOM, Lucent 2623 NA). Pump and probe lasers are polarization adjusted independently for maximum transmission and combined using a 50/50 directional coupler. After passing through the chip, the pump wavelength is filtered further with a dense wavelength division multiplexed (DWDM) optical filter with low insertion loss of 1.5dB and non-adjacent channel isolation over the full passband of 70dB. The mechanical response is recorded using a network analyzer (Omicron Lab – Bode 100).

As mentioned above, the mechanical motion of the beams modulates the effective modal index as shown in the simulated curves in Fig.4b). Here we calculated the effective refractive index variation in dependence of waveguide separation and the beam widths used in the experiments. From the modulation of the refractive index an attractive gradient optical force arises, calculated in Fig.4c) as $\frac{F_{opt}}{PL} = \frac{n_g}{cn_{eff}} \frac{\partial n_{eff}}{\partial g}$, where $n_g$ is the group index, $n_{eff}$ the effective index, $c$ the vacuum speed of light and the partial derivative is carried out with respect to the separation between the beams. The force is normalized to the optical power $P$ and the beam length $L$ and thus given in units of pNµm$^{-1}$mW$^{-1}$. For our device geometry with beams of 350nm width and a waveguide gap of 150nm we expect normalized optical forces up to 5.7pNµm$^{-1}$mW$^{-1}$, comparable to silicon optomechanical devices[4,5,36]. Because of the low mass of the free-standing mechanical beams down to 7.4pg, optical forces on that scale provide an efficient driving mechanism[37]. We therefore easily observe the driven response in both amplitude and phase, as depicted in Fig.4d). Furthermore, by increasing the modulation depth of the modulated pump optical signal, mechanical non-linearities manifest at higher driving amplitude[4,36,12]. In our devices the diamond thin film possesses internal stress which results from temperature gradients during the diamond overgrowth. The thermal expansion mismatch between the diamond film and the SiO$_2$/Si substrate produces compressive stress, when the sample is cooled down from the deposition temperature. The growth process itself produces intrinsic tensile stress which can counteract some of the thermally induced compressive stress[38]. Depending on the location on the wafer from which the chip was fabricated, both compressive and tensile internal stress can therefore be found. The sign of the internal stress directly influences the non-linear behavior of the mechanical beam: for compressive stress a softening Duffing-nonlinearity is expected, whereas for tensile stressed film a stiffening response should occur. This is indeed observed in the strongly driven optomechanical response of our devices. In Fig.4e) we show the mechanical response at increasing driving amplitude for a compressively stressed device. Here a clear softening resonant behavior is observed, in agreement with silicon optomechanical resonators.[4] When on the other hand a tensile stressed beam is actuated, stiffening Duffing-nonlinearity as shown in Fig.4f) results, which is also found in tensile stressed SiN optomechanical resonators[39].

## Discussion

Our chip-scale implementation of diamond opto-mechanical devices provides the ingredients for a flexible platform for functional devices, operational over a wide wavelength range. Contrary to previous beliefs we demonstrate here that polycrystalline CVD diamond layers allow for the fabrication of large-scale nanophotonic circuits without the need for post-growth wafer-bonding techniques. Because CVD diamond offers optical transparency above 220nm all the way into the far infrared wavelength region, diamond is an attractive material for optical applications. Furthermore, because of the large bandgap, excitation of free carriers as in silicon photonic devices does not occur, which is a prerequisite for high power and nonlinear optical applications. In this respect the high thermal conductivity of diamond is equally important, because excess heat can be efficiently distributed away from the waveguide region. This enables us to demonstrate optomechanical interactions in diamond nanocircuits, with high mechanical quality factors and displacement sensitivity. We anticipate that by rigorously scaling our devices towards lengths of a few micrometers, ultra-high mechanical frequencies in the gigahertz range are within reach. This will enable high-frequency sensing applications as well as pique fundamental interest in mechanical coupling to microwave transitions in NV- color centers.

## Methods

**Device fabrication**

Our integrated optical components are realized starting with commercial silicon wafers, single-sided polished. The substrates are thermally oxidized to a target oxide thickness of 2μm. The oxide thickness is optimized in order to provide best coupling efficiency with our grating coupler design. Microcrystalline diamond is then deposited with a thickness of 600nm by plasma enhanced CVD. Ridge waveguides are fabricated using a combination of electron beam lithography and reactive ion etching. In order to provide high etching selectivity we employ Fox15 electron beam photoresist, which is deposited with a thickness of 500nm on the diamond thin film. A timed reactive ion etching process is then performed in oxygen/argon chemistry, with a typical etch rate of 25nm/minute. For the optimal coupling efficiency, the diamond thin film is etched down to a residual thickness of 300nm. For the fabrication of free-standing structures a second lithography step is used to define opening windows for subsequent wet etching. We employ chromium as the masking material against the RIE plasma. The chromium layer is structured using a PMMA positive resist mask. Afterwards the chromium layer in the exposed areas is removed using chromium etchant solution (Sigma Aldrich Chromium Etchant 651826) and the PMMA is subsequently removed in acetone and ashing in oxygen plasma. After the second RIE step, the underlying oxide layer is isotropically etched in buffered oxide etch (6%) and subsequently in chromium etchant to remove the remaining chromium layer throughout the wafer. Finally the finished samples are transferred to methanol and rapidly dried on a hot plate.

**Measurement setup**

For optical characterization, fabricated devices are mounted on a computer controlled three axis motorized stage. The stage is installed inside a vacuum chamber, which allows us to reach a base pressure of $10^{-6}$ mTorr. An optical fiber array with eight single mode fibers is used to launch light from a tunable laser source (Santec TSL 510, New Focus Venturi TLB 6600) into the chip. After transmission through the on-chip devices, light emerging at the output coupler is collected via the optical fiber array and is recorded with a low noise photodetector (New Focus 2011 and New Focus 1811). Electrical analysis of the mechanical motion is carried out using an network / spectrum analyzer (R&S ZVL6). For the driven response, the input optical signal is amplified with an erbium doped fiber amplifier (Pritel LNHPFA-30). Optical modulation is applied prior to amplification using a lithium niobate electro-optical modulator (Lucent 2623NA) and the optomechanical response is recorded using a vector network analyzer (Omicron Lab – Bode 100).

**Surface morphology and scattering loss of microcrystalline diamond thin films**

In order to estimate the contributions of different mechanisms to the propagation loss within the as-grown diamond films the samples were characterized by atomic force microscopy (AFM) and Raman spectroscopy. The surface morphology was determined prior to nanofabrication after the deposition of the diamond thin film onto the oxidized silicon carrier wafers. A typical AFM result is shown in Fig.5a), measured over a scanning region of 5x5 µm². From the data, the microcrystallinity of the sample is clearly visible, as also shown in the cross-sectional SEM in Fig.1c). The typical grain size is on the order of 100nm, leading to a continuous, columnar structure of individual diamond crystals. From the AFM data we determine a root-mean-square (rms) roughness of 15.4 nm. Individual peaks show elevations up to 100nm.

From the measured rms roughness an upper bound for the expected propagation loss due to scattering can be determined. According to Payne and Lacey[32,33], the scattering loss scales quadratically with the surface roughness $\sigma$ according to the following relation

$$\alpha = 4.34 \frac{\sigma^2}{k_0 d^4 n} \cdot \frac{gf}{\sqrt{2}}$$

In the above equation $\alpha$ is the scattering loss in units of dB/unit length, $k_0$ is the free space wave vector, $d$ is the half height of the waveguide and $n$ the refractive index of the waveguiding layer. The analytical function $g$ depends on the waveguide geometry, while the function $f$ provides a dependence on the step index of the waveguide and the correlation length of the surface roughness. Depending on the statistics used to describe the surface roughness, the factor $\kappa = \frac{gf}{\sqrt{2}}$ is bounded and an upper limit for $\alpha$ can be determined. Therefore an upper bound for the scattering loss can be calculated from the AFM rms data as $\alpha \leq 59.5 dB/cm$ ($\alpha \leq 94.2 dB/cm$) for exponential (Gaussian) statistics, which is in good agreement with the measured propagation loss of 53dB/cm.

**Characterization of the diamond thin film with Raman spectroscopy**

A further possible loss source in our waveguides is the inclusion of non-bound carbon or graphite. In order to estimate the content of non-bound carbon within the as-grown films, Raman spectroscopy was performed, which is often applied to characterize nano-, poly- and single-crystalline diamond films. The sample was characterized at room temperature, using an argon-ion laser at 458nm wavelength with a laser spot of 200µm, focused onto the diamond surface with a laser power of 300mW. A measured spectrum is shown in Fig.5b). Generally, these spectra show a prominent Stokes shift at 1332cm$^{-1}$ which has been identified as parameter for high quality single-crystalline diamond with a minimum of scattering intensity at other wavenumbers. Nano- and poly-crystalline CVD diamond films however, show complex bands with scattering intensity reaching from 1100 – 1600cm$^{-1}$ [40]. These lines are assigned to various constituents such as sp2 bonded carbons, diamond progenitors, amorphous carbon, etc.

In our data a well-defined diamond (sp3) Raman peak at 1332cm$^{-1}$ is detected. The full width at half maximum (FWHM) of this peak is about 7cm$^{-1}$ which is typical for such material[41]. The peak at 1150cm$^{-1}$ corresponds to trans-polyacetylene (trans-CHx) which is found in nano-crystalline CVD diamond[42]. The weak peak at 1350cm$^{-1}$ is attributed to the D-band ("disordered

carbon"[43]) and the band around 1510cm$^{-1}$ is due to an overlap of the G-band (graphite) and trans-CHx of diamond at 1480cm$^{-1}$. These data indicate that non-diamond bound carbon is present most likely only in the nucleation layer of the film. From the Raman data we estimate that the relative part of non-sp3 bound carbon will be below 2% and therefore constitute only a minor part of the measured propagation loss.

Additional complications in the interpretation of the Raman spectra arise from the fact that the intensity of scattering from sp2-bonded carbon is very dependent on the excitation wavelength because of resonance effects, and that scattering from diamond is quite dependent on crystallite size as would be expected from a phonon spectrum. The interpretation of Raman spectra for a discussion of sp2 and sp3 contents in diamond films is therefore only possible after careful calibration using near-edge x-ray absorption fine-structure (NEXAFS) experiments, a characterization technique that unequivocally distinguishes between sp2 and sp3 bonded carbon.

From NEXFAS[44] measurements, the structural properties of high quality nano-diamond films have been evaluated to determine the ratio of sp2 to sp3 bonding. The measurements reveal that no more than 1% is sp2. In NEXFAS the relative sensitivity to sp2 and sp3 bonding is roughly the same. In contrast, Raman spectroscopy is 50 – 100 times more sensitive to sp2 bonding. Thus it is reasonable to assume that the actual percentage of sp2 bonding in our high quality polycrystalline diamond is likely less than 1%, although Raman spectra indicate somewhat "poorer quality".

**Acknowledgement**

W.H.P. Pernice acknowledges support by the Emmy-Noether Program through DFG grant PE 1832/1-1 and by the Karlsruhe Institute of Technology (KIT) through the DFG Center for Functional Nanostructures (CFN) within subprojects A6.04. We thank S. Diewald for help with device fabrication und M. Thuermer for helpful discussions on data analysis. We also thank M. Wolfer for performing the Raman measurement.

**Author contributions**

WP perceived the experiment. PR fabricated the devices and performed the measurements with assistance from SK. CW performed the deposition of the diamond thin films. WP, PR and CN analyzed the data. PR, SK, CN, CW and WP discussed the results and wrote the paper.

**Competing Financial Interests**

The authors declare no competing financial interests.

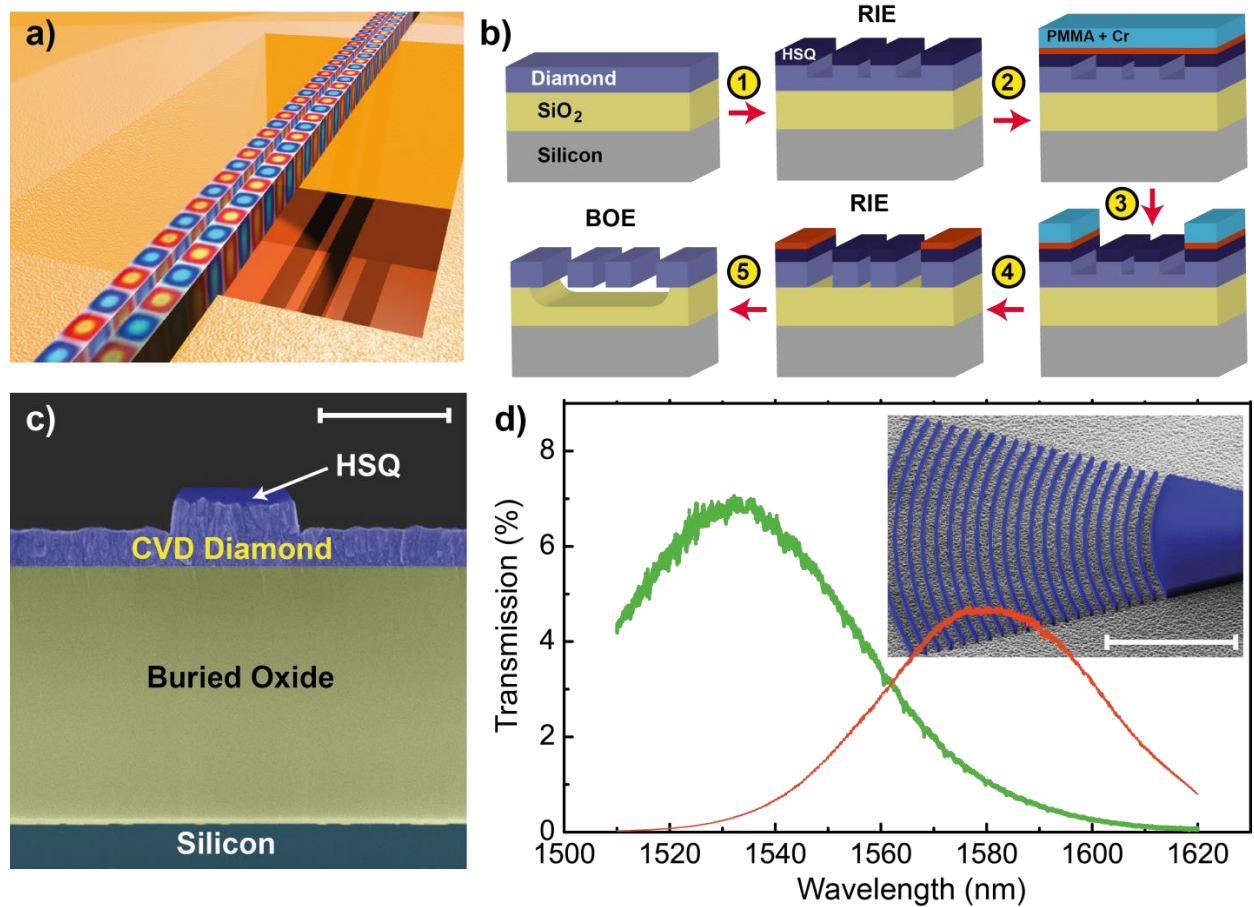

**Figure 1. Diamond optomechanical resonators.** a) Schematic of coupled free-standing waveguides, which act as mechanical resonators. Propagating optical modes are overlaid in color. b) Fabrication routine used to prepare both photonic circuitry and mechanical elements on chip. c) Cross-sectional SEM image of a diamond nanophotonic ridge waveguide. Individual layers are marked in false-color. Scalebar: 1μm. d) Transmission curves of diamond waveguides connected to focusing grating couplers (inset: SEM image of a fabricated device, scalebar: 7.5μm). The central coupling wavelength is tuned by adjusting the period of the grating.

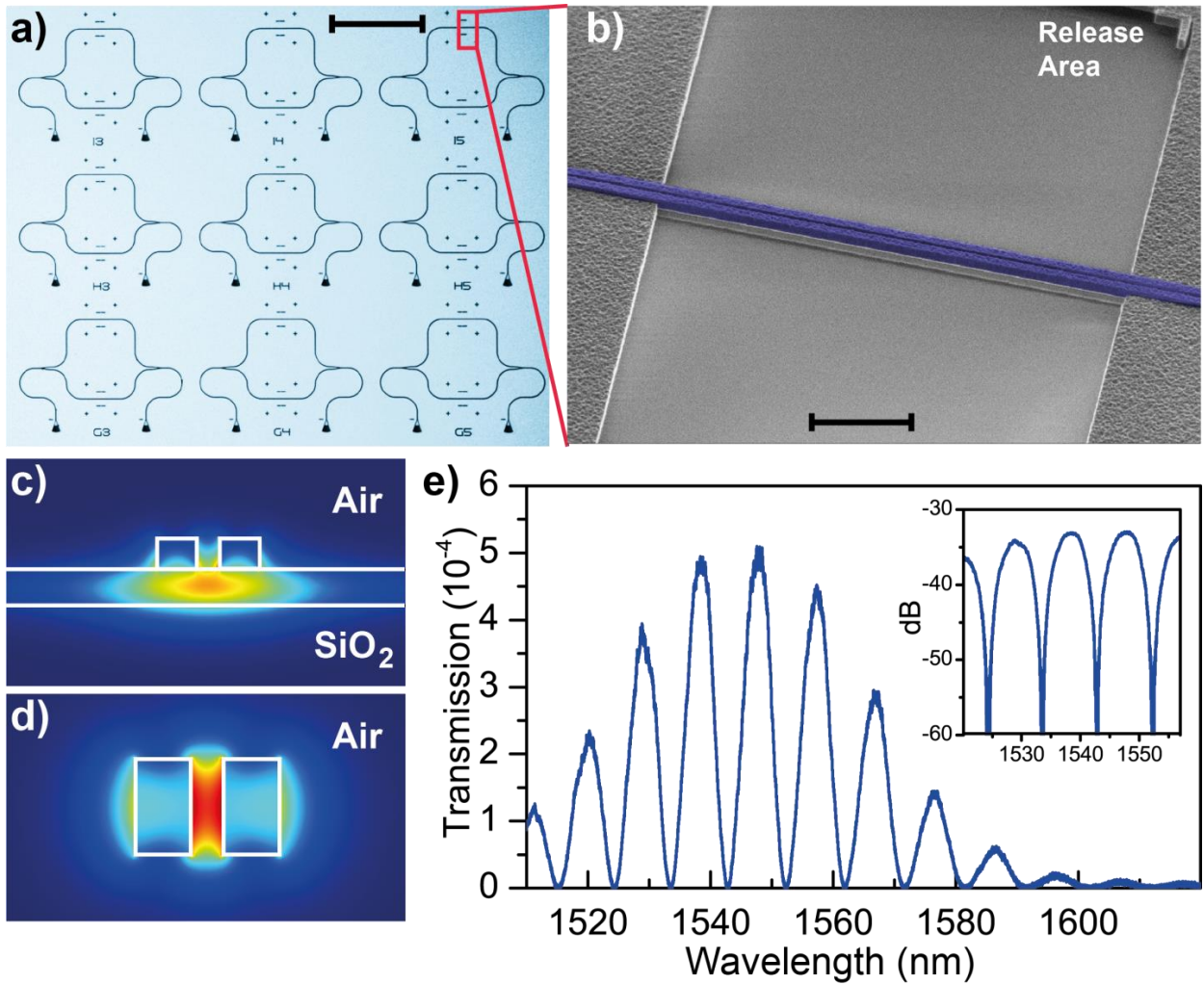

**Figure 2. Diamond integrated nanophotonic circuits.** a) An optical micrograph of a section of a fabricated chip, showing several Mach-Zehnder interferometers connected to grating coupler input/output ports. Scalebar: 250µm. b) SEM image of the released portion of a slotted nanophotonic waveguide. Equivalent waveguide sections are released in both MZI arms. Scalebar: 5µm. c) Simulated mode profile of the unreleased slot waveguide, E-field norm. d) Simulated mode profile of the free-standing diamond waveguide beams, showing strong field concentration in the air gap between the dielectric waveguides. e) Measured transmission profile of a fabricated MZI device with a free-spectral range of 9.5nm. Inset: logarithmic plot of the transmission, showing extinction ratio up to 25dB.

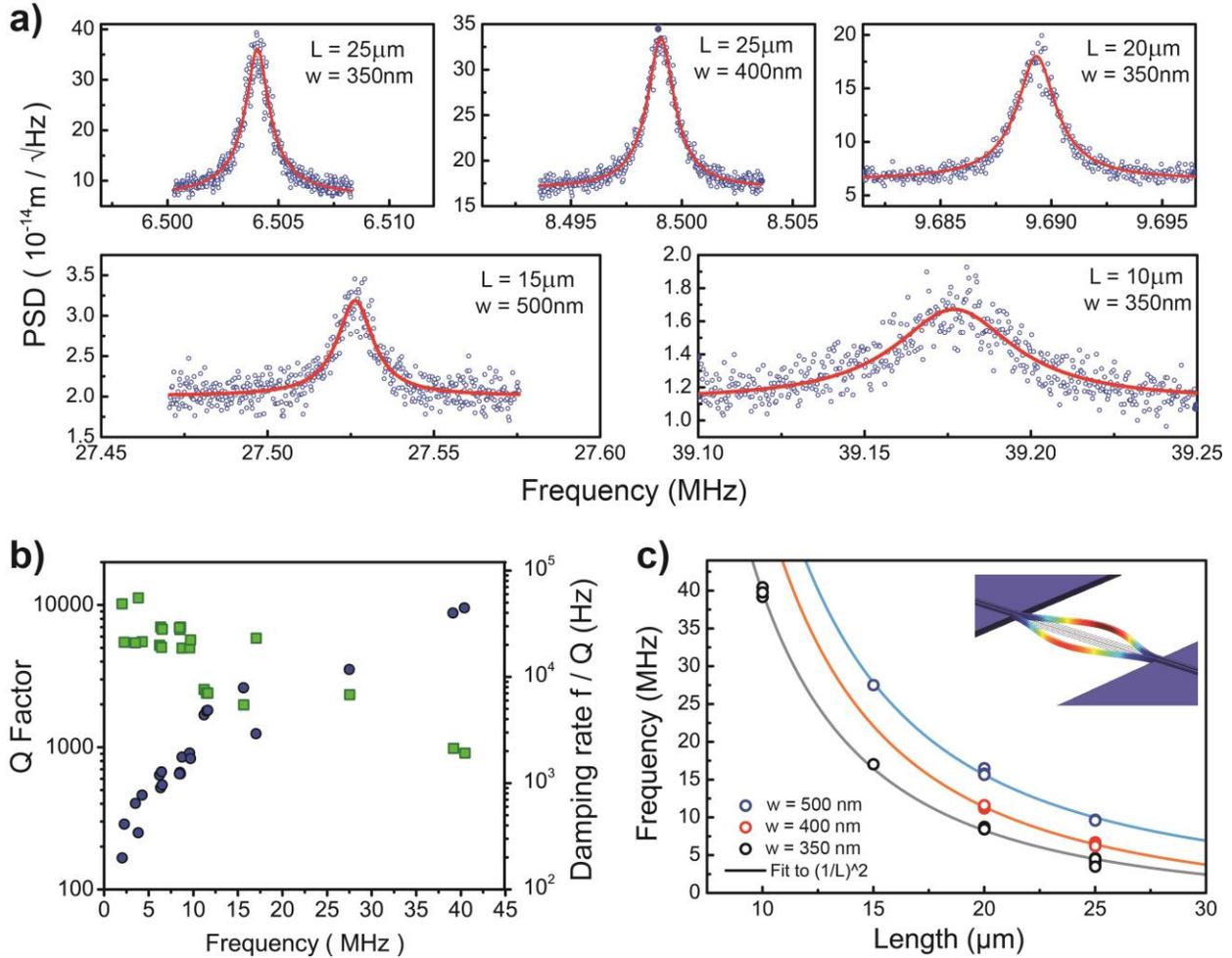

**Figure 3. Thermomechanical motion of diamond resonators.** a) Measured thermomechanical noise spectra of diamond resonators with varying length and width. The designed resonator dimensions are included in the subpanels. b) The extracted mechanical quality factors of fabricated devices in dependence of frequency (green markers). Best Q-factors of 11,200 are obtained at a frequency of 3.8MHz. Corresponding damping rates f/Q are plotted on the right axis (blue markers) c) The measured dependence of resonance frequency on beam length for different beam cross-sections. The shortest devices (10µm length) reach frequencies up to 40MHz. Inset: Simulated mechanical mode of the slot-waveguide geometry.

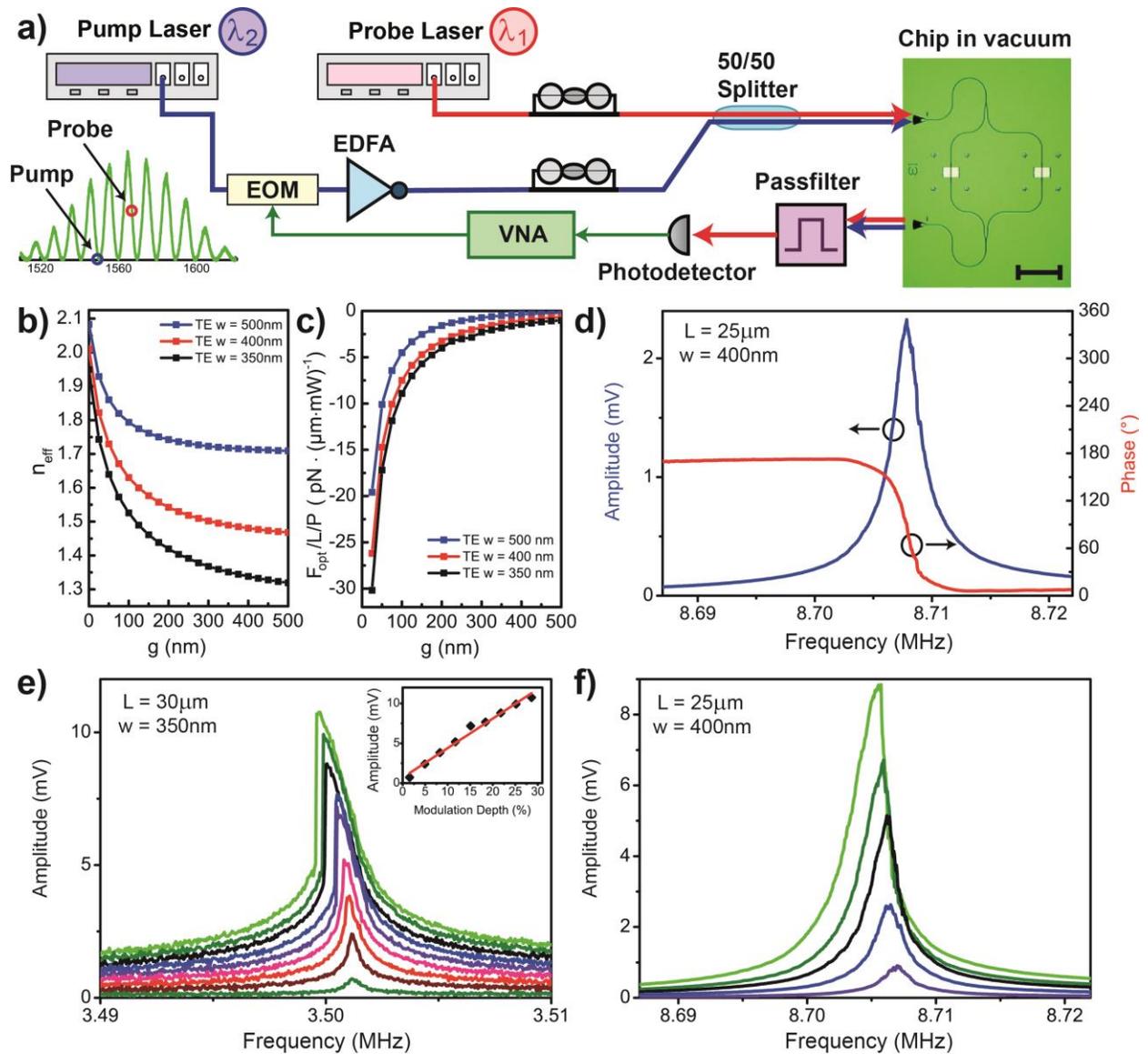

**Figure 4. Driven optomechanical response.** a) Schematic of the pump-probe measurement setup used to characterize the mechanical response. Scalebar in the optical micrograph: 100μm. b) The calculated dependence of the modal refractive index on the slot waveguide geometry. (c) Derived optical forces normalized to beam length $L$ and optical power $P$ are on the order of pNμm$^{-1}$mW$^{-1}$. d) The measured mechanical response in both amplitude (blue) and phase (red) for a sample device at 8.7MHz. e) Duffing softening non-linearity in compressively stressed optomechanical devices for increasing modulation depth from 1.5% to 28.6%. f) Corresponding

stiffening Duffing non-linearity in tensile stressed nanomechanical resonators, for a modulation depth between 8.5% and 74%.

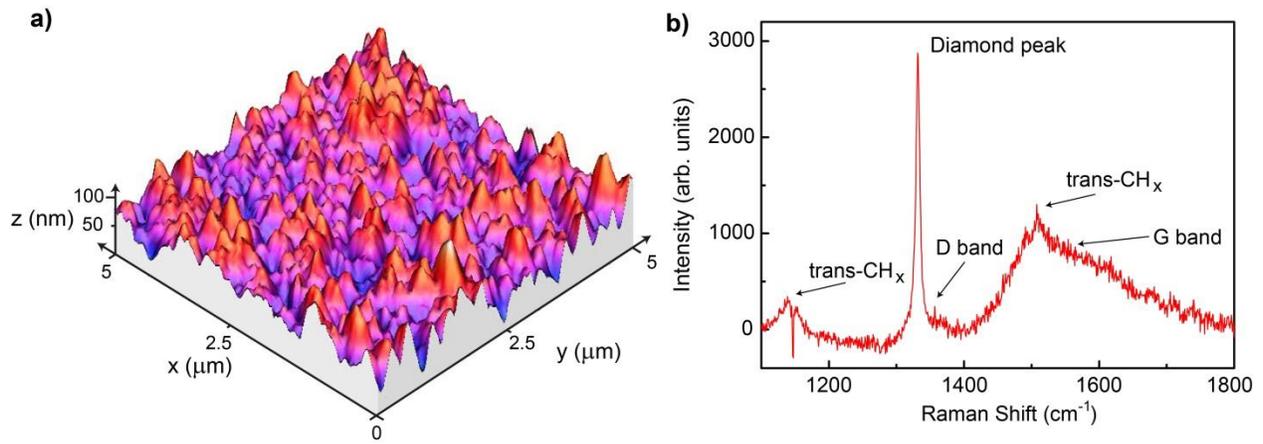

**Figure 5. Characterization the microcrystalline diamond thin film.** a) Atomic force microscope (AFM) measurement of the surface morphology of a diamond thin film before lithographic structuring. Root-mean-square roughness of 15.4nm is obtained. b) Measured Raman spectrum of a microcrystalline diamond thin film. The data shows the results of Raman spectroscopy at room temperature of an as-grown diamond thin film, with a strong diamond peak at 1332cm$^{-1}$.